\documentclass[twocolumn,         
               showpacs,            
               preprintnumbers,     
               aps,                 
               prd,          	    
               a4paper,             
               superscriptaddress,      
               nofootinbib,         
               tightenlines,        
               floats,floatfix,11pt
               ]{revtex4-2}              
\usepackage{graphicx}  
\usepackage[margin=0.6in]{geometry}
\usepackage{dcolumn}   
\usepackage{bm}  
\usepackage[normalem]{ulem}
\usepackage[utf8]{inputenc}
\usepackage{amsmath,amssymb}
\usepackage{soul}
\usepackage{float}
\usepackage[thinlines]{easytable}
\usepackage{array,booktabs}
\usepackage{lipsum} 
\usepackage{multirow}
\usepackage{placeins} 
\usepackage[colorlinks=true,linkcolor=blue,citecolor=blue]{hyperref}
\usepackage{subcaption}
\usepackage{array,makecell}
\usepackage{cleveref} 
\begin{document} 

\title{Constraint on Momentum-coupled Dark Energy using DESI DR2 }
\author{Prasanta Sahoo }
\email{prasantmath123@yahoo.com} 
\affiliation{Midnapore College (Autonomous), Midnapore, West Bengal, India, 721101}
\author{Nandan Roy}
\email{nandan.roy@mahidol.ac.th (Corresponding Author)} 
\affiliation{NAS, Centre for Theoretical Physics \& Natural Philosophy, Mahidol University,
Nakhonsawan Campus, Phayuha Khiri, Nakhonsawan 60130, Thailand}
\author{Himadri Shekhar Mondal }
\email{himumath100@gmail.com} 
\affiliation{Midnapore College (Autonomous), Midnapore, West Bengal, India, 721101}


\begin{abstract}
In this work, we study two scalar field–driven dark energy models characterized by the axion potential and the inverse power-law potential, each coupled to dark matter through a momentum-exchange interaction. By formulating the dynamics as an autonomous system, we identify the equilibrium points and analyze their stability. To constrain these models, we utilize observational data from Pantheon Plus Type Ia Supernovae, DES Y5, DESI DR2 BAO, and Planck 2018 CMB compressed likelihood, employing Markov Chain Monte Carlo (MCMC) methods. Both potentials exhibit weak to strong preference over the $\Lambda$CDM model, with a particularly strong preference for the momentum-coupled scenario when Supernova data are included in the analysis. Furthermore, we find the coupling parameter to be negative, with no lower bound, for both potentials. This suggests that momentum-exchange coupling between the dark sectors cannot be ruled out. From the stability analysis, we observe that for both potentials, the late-time attractor corresponds to a dark energy–dominated phase, and the scalar field can behave as a stiff fluid during the early epoch.
\end{abstract}

\maketitle

\section{Introduction}

With the increase in our precision of cosmological measurements, the deviation from the standard model of cosmology, which is generally referred to as $\Lambda$CDM, is becoming more prominent.  Recent findings from the Dark Energy Spectroscopic Instrument (DESI) \cite{DESI:2024mwx, DESI:2024aqx, DESI:2025zgx,DESI:2024jis,DESI:2025fii}  and the Year 5 supernovae sample of the Dark Energy Survey (DES)\cite{DES:2024jxu} suggests that at low redshift, the BAO and supernovae data might not prefer the standard $\Lambda$CDM model whereas the dark energy might be dynamical in nature.

Moreover, analyzing different cosmological parameters through various data combinations reveals statistically significant discrepancies. The most prominent of these is the Hubble tension, which reflects a statistically notable discrepancy of nearly $\simeq 5.3 \sigma$~\cite{Riess_2022} between early-universe-derived measurements and those from the late-time distance ladder concerning the current Hubble parameter ($H_0$). Estimates utilizing early universe data, such as the CMB from the Planck collaboration ~\cite{Planck2020}, BAO ~\cite{BAO2017, BAO2011, DES2018,DES:2018rjw}, and BBN ~\cite{BBN2021}, suggest the Hubble constant at $H_0 \sim (67.0 - 68.5)$ km/s/Mpc with $\Lambda$ as the dark energy component. Conversely, late-time distance ladder techniques, including the SH0ES project ~\cite{Riess:2021jrx} and H0LiCOW collaborations ~\cite{Wong:2019kwg}, have found $H_0$ to be $H_0 = (73.04 \pm 1.42)$ km/s/Mpc and $H_0 = (73.3_{-1.8}^{+1.7}$ km/s/Mpc, respectively. The potential dynamical nature of DE is more preferable in light of these findings and encourages the search for alternatives to the $\Lambda$CDM model.

The simplest model of dynamic dark energy can be constructed by treating dark energy as a fluid, in which various cosmological variables like the Hubble Parameter, the equation of state, and the density parameters of dark energy are parameterized. Alternatively, scalar field models, such as quintessence, k-essence, and phantom dark energy, have been suggested \cite{amendola2010dark, Bamba:2012cp}. These models typically attribute the universe's accelerated expansion to a scalar field moving through a potential, generating an effective negative pressure \cite{copeland2006dynamics, Peebles2003, Armendariz2001, roy2022quintessence, Banerjee:2020xcn, Lee:2022cyh, Krishnan:2020vaf}. While dark energy is usually presumed to evolve without non-gravitational interactions, whether it interacts with dark matter remains unresolved. Interacting dark energy (IDE) models were proposed to tackle the cosmic coincidence problem \cite{Cai:2004dk,mangano2003coupled,Sadjadi:2006qp,Wang:2016lxa, Jesus:2020tby}. In IDE models, dark matter and dark energy densities are interconnected via an interaction term that facilitates energy and/or momentum transfer, affecting cosmic evolution \cite{Wang:2016lxa}. Numerous studies investigate how these interactions influence cosmological measurements \cite{amendola2000coupled, farrar2004interacting, mangano2003coupled, tamanini2015phenomenological, chimento2010linear, pan2015analytic, pettorino2005extended, pettorino2008coupled}. Recently, these models have shown potential to alleviate the $H_0$ and $\sigma_8$ tensions \cite{Salvatelli2014,Costa2017,DiValentino2019,Kumar2020,DiValentino:2019ffd,DiValentino:2019jae,Yang:2018euj,Wang:2018duq}. Both cosmological observations and dynamical systems analysis have been used to examine the IDE models' dynamics \cite{Salvatelli2014,Costa2017,DiValentino2019,Kumar2020,DiValentino:2019ffd,DiValentino:2019jae,Yang:2018euj,Wang:2018duq,Khyllep:2021wjd,Caldera-Cabral:2008yyo,Amendola:1999er,Boehmer:2008av,Zonunmawia:2017ofc,Hussain:2022dhp,Bahamonde:2017ize}. 

In the energy transfer IDE models, dark matter and dark energy are jointly conserved rather than independently, and they are connected via an interaction term.  Each interaction form generates distinct phenomenological imprints and cosmic evolution \cite{Roy:2018eug,Roy:2023uhc,Kritpetch:2024rgi}. However, these models do not have a Lagrangian description, and their interaction term, which is inherently ad hoc, is incorporated at the equation level. In \cite{Pourtsidou:2013nha}, the authors applied the pull-back formalism to fluids to extend the fluid action, integrating interactions between dark energy, modelled by a scalar field, and dark matter. This approach resulted in three different coupled model families. Types 1 and 2 include both energy and momentum exchanges between dark energy and dark matter. Whereas, Type 3 models correspond to pure momentum transfer interactions. Some previous coupled quintessence models have examined the direct momentum exchange between dark matter and dark energy \cite{Pourtsidou:2016ico,Chamings:2019kcl,BeltranJimenez:2021wbq,Liu:2023rvo,Koyama:2009gd,Caldera-Cabral:2009hoy}. For example, the effect of a coupled quintessence model with pure momentum exchange on the structure formation has been studied in \cite{Pourtsidou:2016ico,Chamings:2019kcl}. Observational constraints on these models have been reported in \cite{BeltranJimenez:2021wbq}. This type of coupling has been studied in the setup of the early dark energy (EDE) scenario in \cite{Liu:2023rvo}. Although the effects of these momentum-coupled dark energy models on the background and linear perturbations have been studied, a dynamical system analysis of these models remains unexplored. 

In this study, we employ dynamical systems analysis to investigate a model of pure momentum transfer \cite{Pourtsidou:2016ico}, using a dynamical framework in conjunction with current cosmological observations. Dynamical systems analysis offers a qualitative approach to studying non-linear systems and is widely used to assess the stability and long-term behavior of interacting dark energy models \cite{Khyllep:2021wjd, Caldera-Cabral:2008yyo, Amendola:1999er, Boehmer:2008av, Zonunmawia:2017ofc, Hussain:2022dhp, Bahamonde:2017ize}. To constrain the cosmological parameters, we utilize observational data from the Pantheon+ Type Ia Supernovae compilation, DES Y5, DESI DR2 BAO, and the compressed Planck 2018 CMB likelihood. Additionally, we evaluate model selection criteria by computing the Akaike Information Criterion (AIC).

The manuscript is structured as follows. In Section II, we present the mathematical framework for formulating the momentum transfer model. Section III introduces the dynamical system in terms of polar variables and defines the system for both the axion potential and the inverse power-law potential. Section IV outlines the observational data sets used in the analysis. Section V details the constraints obtained through Markov Chain Monte Carlo (MCMC) analysis. Section VI discusses the statistical criteria employed for model selection, while Section VII examines the evaluation of key cosmological parameters. In Section VIII, we analyze the stability of the model. Finally, Section IX summarizes our findings and presents the conclusions.

\section{Mathematical Background}
In a FLRW universe where the geometry is flat, the metric can be represented by
 \begin{equation}\label{line_element_momentum}
 ds^{2}=-dt^{2} + a^{2}(t)dx^{i}dx^{j};\quad i,j=1,2,3;
 \end{equation}
 Here $a(t)$ is the scale factor of the universe, and the components of the universe are standard matter and a scalar field $\phi$ in the form of quintessence representing dark energy. Motivated by the fact that an interaction in dark sector may impart a transfer of momentum, we consider a pure momentum transfer (or Momentum coupling) term $\mathcal{L}_{int}=-\beta (u^{\alpha}\partial_{\alpha}\phi)^{2}$ in terms of the 4-vector $u^{\mu}=(1, 0, 0, 0)$ with $\beta$ as a coupling constant. Then the total Lagrangian is defined as 
\begin{equation}\label{Lagrangian}
\mathcal{L}=  -\frac{1}{2}\partial^{\alpha}\phi \partial_{\alpha}\phi -V(\phi) + \mathcal{L}_{int},
\end{equation}
where $V(\phi)$ is the potential of the scalar field. Then the action for such a universe is given by
\begin{equation}\label{action}
S= \int d^{4}x\sqrt{-g}\mathcal{L}.
\end{equation}
The variation of the action (\ref{action}) with respect to the metric $g^{\mu \nu}$ gives the energy-momentum tensor   
\begin{equation}
    T_{\mu\nu}= \partial_{\mu}\phi \partial_{\nu}\phi +2\beta (u^{\alpha}\partial_{\alpha}\phi)u_{\mu}u_{\nu} + g_{\mu \nu} \mathcal{L} .
\end{equation}

Then the energy density and pressure of the scalar field are respectively given by
\begin{equation}\label{scfenergy_eq_momentum}
\rho_{\phi}=T_{00}=\frac{1}{2}(1-2\beta)\dot{\phi}^{2}+V(\phi) , 
\end{equation}

\begin{equation}\label{scfpressure_eq_momentum}
p_{\phi}=\frac{1}{3}\displaystyle\sum_{i=1}^{3}T_{ii}=\frac{1}{2}(1-2\beta)\dot{\phi}^{2}-V(\phi).
\end{equation}

The Friedmann equations for this model are  given by
 \begin{equation}\label{H_eq_momentum}
 H^{2}=\frac{\kappa^{2}}{3}\left(\rho_{m}+\frac{1}{2}(1-2\beta)\dot{\phi}^{2}+V(\phi)\right) ,
 \end{equation}

 \begin{equation}\label{Hdot_eq_momentum}
\dot{H}=-\frac{\kappa^{2}}{2}\left( \rho_{m}+ (1-2\beta)\dot{\phi}^{2}\right) .
\end{equation}

Here $\kappa^{2}=8\pi G$, $\rho_{m}$ is the energy density of matter, $p_{m}=0$ is the pressure of matter, $H=\dot{a}/a$ is the Hubble parameter, and the over-dot represents derivative with respect to cosmic time $t$. One can write that the corresponding Klein-Gordon equation becomes

\begin{equation}\label{KG_eq_momentum}
\ddot{\phi}+3H\dot{\phi}+\frac{1}{(1-2\beta)}\frac{dV(\phi)}{d\phi}=0 .
\end{equation}

The continuity equations for matter and scalar field are given by
\begin{equation}\label{Continuity_eq_momentum}
 \dot{\rho_{i}}=-3H(p_{i}+\rho_{i}),\quad \forall \quad i = m, \phi .
\end{equation}

To ensure that the model is physically viable, we consider $\beta<\frac{1}{2}$. This kind of coupling faces strong coupling issues for $\beta=\frac{1}{2}$ and $\beta>\frac{1}{2}$ gives the negative kinetic term, which indicates the appearance of ghost in the theory \cite{PhysRevD.94.043518}.
The expression for density parameter for a given species `i' is given by $\Omega_{i}=\frac{k^{2}\rho_{i}}{3H^{2}}$. Accordingly, the Friedmann constraint becomes $\Omega_{m}+\Omega_{\phi}=1$. Since we are interested in the late-time dynamics of this interaction, we have neglected radiation from the total energy budget.

\section{The Dynamical System}
 For the better mathematical handling of the system of equations and to perform stability analysis, we introduce the following dimensionless transformations: 
\begin{align}\label{transformation_polar}
 r \cos \theta & = \frac{\kappa \dot{\phi}}{\sqrt{6}H}, \quad r \sin \theta= \frac{\kappa \sqrt{V (\phi)}}{\sqrt{3}H}, \\ \nonumber
 \lambda &= -\frac{1}{k V}\frac{dV (\phi)}{d\phi}, \Gamma =\frac{V \frac{d^{2}V}{d\phi^{2}}}{\left( \frac{dV}{d\phi} \right)^{2}}.
\end{align}

Here $\lambda$ and $\Gamma$ are the slope and tracking parameters, respectively. They are essential to close the above dynamical system for some particular choice of the scalar field potential. A similar polar transformation has been previously used in \cite{Roy:2013wqa,Roy:2018nce,Sahoo:2025vtu}, and it has been shown that writing the system of equations in polar form is very useful for the numerical investigation of the system since the polar variables can be directly related to the cosmological observables. With these transformations, the system of equations transforms to the following,

\begin{subequations}\label{polar_autonomous:1_momentum}
\begin{align}
    r^{\prime} &= \frac{3r }{2}\left( 
r^{2}(\cos\hspace{0.02cm}2\theta -2\beta \cos^{2}\theta) - \cos\hspace{0.02cm} 2\theta \right) \nonumber \\
  &\quad+ \frac{\sqrt{6}\beta}{(1 - 2\beta)}\lambda r^{2} \cos\hspace{0.02cm}\theta \sin^{2}\theta,\label{eq:r_momentum} \\
    \theta^{\prime} &= \frac{3}{2}\sin\hspace{0.02cm} 2\theta -\sqrt{\frac{3}{2}}\lambda r \sin\hspace{0.02cm}\theta \left( 
\cos^{2}\theta + \frac{\sin^{2}\theta}{1-2\beta} \right),\label{eq:theta_momentum} \\
    \lambda^{\prime} & = -\sqrt{6} r \hspace{0.02cm}\cos \theta \left(\Gamma - 1 \right)  .\label{eq:lam_momentum}
\end{align}
\end{subequations}

A prime denotes the derivative with respect to the e-fold parameter $N=ln(a/a_{0})$, where $a_{0}$ is the present expansion rate of the universe. In this dimensionless setup, the Friedmann constraint, the equation of state (EoS) of the scalar field, and the total equation of state are respectively given in Eq.(\ref{Friedmann_constraint_polar})- Eq.(\ref{Total_EoS}). 

\begin{equation}\label{Friedmann_constraint_polar}
 r^{2}(1-\beta - \beta \hspace{0.04cm}cos\hspace{0.04cm}2\theta)+\Omega_{m}=1,
\end{equation}

\begin{equation}\label{EoS_quintessence}
 w_{\phi}=\frac{(1-2\beta)cos^{2}\theta - sin^{2}\theta}{(1-2\beta)cos^{2}\theta + sin^{2}\theta},
\end{equation}

\begin{equation}\label{Total_EoS}
 \omega_{tot}= r^2 \cos 2\theta -2\beta r^2 \cos^{2}\theta.
\end{equation}

\begin{equation}
    q=-1+\frac{3}{2}r^2(\cos(2\theta)-2\beta \cos\theta^2) 
\end{equation}

As $\lambda$ and $\Gamma$ in Eq.(\ref{transformation_polar}) both depend on the choice of potential of the scalar field, the equation Eq.(\ref{eq:lam_momentum}) also depends on the choice of the potential. To close the system of equations Eq.(\ref{polar_autonomous:1_momentum}), one needs to choose a particular form of the potential and find out the corresponding $\Gamma$, preferably represented as a function of $\lambda$ and substitute it in equation Eq.(\ref{eq:lam_momentum}). For viable late-time dynamics of the proposed model, we consider two such potentials of the scalar field: One is the axion potential \cite{Oikonomou:2024ofp, Cheek:2024ofn, LinaresCedeno:2020dte}, and the inverse power law potential \cite{Kneller:2002zh, Lu:2013roa, Riazuelo:2000fc}.

\subsection{The Axion Potential}
The axion potential is represented in the form of two energy scales, namely the axion mass $m_{a}$ and the effective energy scale $f_{a}$, also known as the decay constant. Here we consider the particular form of this potential,
\begin{equation}\label{Axion_potential}
    V(\phi)=m_{a}^{2}f_{a}^{2}\left( 1 + \cos (\phi /f_{a}) \right)^{n},
\end{equation}
where $n$ is a positive constant. For this potential $\lambda=n\sqrt{\alpha}\tan\left( \frac{\phi}{2f_{a}} \right)$ and $\Gamma=1-\frac{1}{2n}-\frac{n\alpha}{2\lambda^{2}}$, where $\alpha=\frac{1}{\kappa^{2}f_{a}^{2}}$. For this potential the system of equations Eq.(\ref{polar_autonomous:1_momentum}) reduces to

\begin{subequations}\label{axion_autonomous:1_momentum}
\begin{align}
    r^{\prime} &= \frac{3r }{2}\left( 
r^{2}(\cos\hspace{0.02cm}2\theta -2\beta \cos^{2}\theta) - \cos\hspace{0.02cm} 2\theta \right) \nonumber \\
  &\quad+ \frac{\sqrt{6}\beta}{(1 - 2\beta)}\lambda r^{2} \cos\hspace{0.02cm}\theta \sin^{2}\theta,\label{eq:r_momentum_axion} \\
    \theta^{\prime} &= \frac{3}{2}\sin\hspace{0.02cm} 2\theta -\sqrt{\frac{3}{2}}\lambda r \sin\hspace{0.02cm}\theta \left( 
\cos^{2}\theta + \frac{\sin^{2}\theta}{1-2\beta} \right),\label{eq:theta_momentum_axion} \\
    \lambda^{\prime} & = \frac{\sqrt{3}}{\sqrt{2}n} r \cos \theta \left( \lambda^{2} + n^{2}\alpha \right);\label{eq:lam_momentum_axion}
\end{align}
\end{subequations}

where $\alpha =\frac{1}{\kappa^{2}f_{a}^{2}}$.

\subsection{The Inverse Power Law Potential}
Here we consider the following form of the inverse power law potential
\begin{equation}\label{InvPow_potential}
    V(\phi)=M^{m+4}\phi^{-m},
\end{equation}
where $M$ is the mass scale of the potential and $m$ is a positive constant representing the steepness of the potential. For $m=0$, the model reduces to a standard $\Lambda$CDM model. For this potential $\lambda=\frac{m}{\kappa \phi}$ and $\Gamma=1+\frac{1}{m}$. Then the system of equations in Eq.(\ref{polar_autonomous:1_momentum}) reduces to

\begin{subequations}\label{invpow_autonomous:1_momentum}
\begin{align}
    r^{\prime} &= \frac{3r }{2}\left( 
r^{2}(\cos\hspace{0.02cm}2\theta -2\beta \cos^{2}\theta) - \cos\hspace{0.02cm} 2\theta \right) \nonumber \\
  &\quad+ \frac{\sqrt{6}\beta}{(1 - 2\beta)}\lambda r^{2} \cos\hspace{0.02cm}\theta \sin^{2}\theta,\label{eq:r_momentum_invpow} \\
    \theta^{\prime} &= \frac{3}{2}\sin\hspace{0.02cm} 2\theta -\sqrt{\frac{3}{2}}\lambda r \sin\hspace{0.02cm}\theta \left( 
\cos^{2}\theta + \frac{\sin^{2}\theta}{1-2\beta} \right),\label{eq:theta_momentum_invpow} \\
    \lambda^{\prime} & = -\frac{\sqrt{6}}{m}r \lambda^{2}\cos \theta.\label{eq:lam_momentum_invpow}
\end{align}
\end{subequations}

We have used the autonomous systems obtained in Eq.(\ref{axion_autonomous:1_momentum}) for the axion and Eq.(\ref{invpow_autonomous:1_momentum}) for the power-law potential to apply constraints to the model using recent cosmological data and to perform a dynamical systems analysis.

\section{Observational Data}

To evaluate how well this model aligns with current cosmological observations, we conducted an MCMC analysis to constrain the main cosmological parameters along with the model-specific parameters. For this purpose, Eq.(\ref{axion_autonomous:1_momentum}) and Eq.(\ref{invpow_autonomous:1_momentum}) have been reformulated as derivatives in terms of the redshift ($z$). We have also utilised the acceleration equation in writing the following form for the numerical integration.

\begin{equation}
    \frac{dH}{dz} = -\frac{3}{2} H^2 (\Omega_m + 2(\beta -1) r^2 Cos\theta^2) /(\frac{dz}{dt})
\end{equation}

The initial condition is set at the present epoch ($z=0$). With this configuration, we can directly calculate the Hubble parameter ($H$) relative to redshift ($z$), which is easier to compare with observed data. Cosmological data analyses using a similar dynamical system setup has been done in \cite{Myrzakulov:2025jpk,Hussain:2024yee}. A more extensive analysis using the dynamical system approach along with the Boltzmann code CLASS is discussed in \cite{Roy:2018nce,Roy:2018eug,Urena-Lopez:2020npg, LinaresCedeno:2021aqk,Roy:2023uhc}. In this work, we performed an MCMC analysis employing the \textit{emcee} library and used the publicly available GetDist plotting code to visualise the posterior distributions. Below, we provide specifics about the dataset used in our analysis.

\subsection{Supernova Data}
Type Ia supernovae are widely used as standard candles because of their comparatively uniform absolute luminosity~\cite{reiss1998supernova,SupernovaSearchTeam:1998fmf}. In the present research, we made use of the Pantheon Plus compilation of SN-Ia data~\cite{Scolnic:2021amr, Riess_2022, Brout:2022vxf} alongside the DES Year 5 data \cite{DES:2024jxu}. The Pantheon+ dataset comprises 1701 light curves with redshifts spanning from $0.01$ to $2.26$. DES-Y5 (Dark Energy Survey, through their Year 5 data release) has recently provided outcomes from a newly and uniformly chosen collection of 1635 photometrically categorized SN Ia within the redshift range of $0.1 < z < 1.3$. This is further supported by 194 low-redshift SN Ia, which are also part of the Pantheon+ sample, covering redshifts from $0.025 < z < 0.1$.

\subsection{DESI BAO Data}

The density of visible baryonic matter exhibits recurring, periodic fluctuations called baryon acoustic oscillations. These oscillations are essential standard rulers for precise distance measurements in cosmology. In this study, we used the 2025 BAO observation data from the Dark Energy Spectroscopic Instrument (DESI-DR2) as noted in the reference ~\cite{DESI:2025zgx, DESI:2024mwx}.  BAO provides measurements of the effective distance along the line of sight as

\begin{equation}
\frac{D_H(z)}{r_d} =\frac{c r_d^{-1}}{H(z)}
\end{equation}

and along the transverse line of sight as,
\begin{equation}
    \frac{D_{M}(z)}{r_{d}}\equiv\frac{c}{r_{d}}\int_{0}^{z}\frac{d\tilde{z}}{H(\tilde{z})}=\frac{c}{H_{0}r_{d}}\int_{0}^{z}\frac{d\tilde{z}}{h(\tilde{z})}.
\end{equation}
The angle average distance is measured as,

\begin{equation}
\frac{D_V(z)}{r_d} =\left[\frac{c z r_d^{-3} d_L^2(z)}{H(z)(1+z)^2}\right]^{\frac{1}{3}}.
\end{equation}

Here, the luminosity distance is represented by $d_L(z)$. In this work, we have considered the sound horizon $r_d$ to be a free parameter to be fitted from the data.

\subsection{Compressed CMB Likelihood}

We have also used the Planck18 compressed likelihood following the approach in ~\cite{Chen:2018dbv} (hereafter CMB).  CMB compressed prior is generally used as an alternative to the global fitting of the Full Planck data set, particularly for dark energy models that extend beyond the standard $\Lambda$CDM. The compressed prior has similar constraining power as the full Planck data.  The compressed likelihood takes into account the baryon physical density $\omega_b = \Omega_b h^2$, as well as two shift parameters: $l_{\mathrm{A}}=\left(1+z_{*}\right) \frac{\pi D_{\mathrm{A}}\left(z_{*}\right)}{r_{s}\left(z_{*}\right)}$ and $\mathcal{R}=\sqrt{\Omega_M H_0^2} D_A\left(z_{*}\right)$, with $z_{*}$ representing the redshift at decoupling and $D_A$ denoting the comoving angular diameter distance. It is important to acknowledge that these priors may not fully capture the complexity of the entire CMB power spectrum due to underlying assumptions, particularly if there are significant deviations from the $\Lambda CDM$ model. Such deviations might lead to systematic biases in cosmological parameters \cite{Corasaniti:2007rf, Elgaroy:2007bv}.

We examined the following three combinations of data sets to constrain the model presented here.

\begin{enumerate}
    \item Set 1: Pantheon Plus + CMB + DESI DR2
    \item  Set 2: DES Y5 + CMB + DESI DR2
    \item Set 3: CMB + DESI DR2
\end{enumerate}

\section{Constraints from Data Analysis}

In this section, we present the observational constraints obtained on the $\Lambda$CDM model and the interacting model for the axion potential and the Inverse Power Law (InvPow) potential using the above three cosmological datasets. In our notation $\Omega_{m}=\Omega_{b}+\Omega_{c}$, where $\Omega_{b}$ and $\Omega_{c}$ are the density parameters for baryons and cold dark matter (CDM), respectively. To constrain our models, we have considered flat priors on the base cosmological parameters $H_0:[60,90], \Omega_{m0}:[0.2,0.4], r_{d}:[120, 160], \Omega_{b}h^{2}:[0.01,0.03]$ for both potentials. For the axion potential, our choice for the model parameters is $\theta_0:[1.22,1.91], \beta:[-1.5, 0.49],\alpha:[0, 10], \lambda_0:[0, 3.5]$, n:[0.5, 3.2], and for the inverse power law potential $\theta_0:[\pi /2 - 0.25, \pi /2 + 0.25], \beta:[-1, 0.49], \lambda_0:[0.5, 6]$, m:[1, 5]. From Eq.(\ref{EoS_quintessence}), it is evident that the scalar field's EoS parameter is influenced by $\beta$ and $\theta$. We have selected a prior for $\beta$ and $\theta$ with the condition that $w_\phi \simeq -1 \pm 0.4$. For further details on the choice of priors on $\theta$ and $\beta$, we refer to Appendix \ref{Prior on model parameters}.

\begin{table*}[ht]
\centering
\resizebox{\textwidth}{!}{%
\begin{tabular}{|l|ccc|ccc|ccc|}
\toprule
Parameters & \multicolumn{3}{c}{Set 1: Pantheon Plus+CMB+DESI DR2} & \multicolumn{3}{c}{Set 2: DES Y5+CMB+DESI DR2} & \multicolumn{3}{c}{Set 3: CMB+DESI DR2} \\
\cmidrule(lr){2-4}\cmidrule(lr){5-7}\cmidrule(lr){8-10}
& $\Lambda$CDM & Axion & InvPow & $\Lambda$CDM & Axion & InvPow & $\Lambda$CDM & Axion & InvPow \\
\midrule
$\Omega_{m0}$ 
    & $0.3178\pm 0.0058$ 
    & $0.3147\pm 0.0064$
    & $0.3140\pm 0.0064$ 
    & $0.3195\pm 0.0058$
    & $0.3156\pm 0.0065$ 
    & $0.3147\pm 0.0063$ 
    & $0.3129\pm 0.0060$ 
    & $0.3170\pm 0.0073$ 
    & $0.3177\pm 0.0072$\\
$H_0$ 
    & $67.36\pm 0.42$ 
    & $-67.60^{+0.44}_{-0.49}$
    & $67.65\pm 0.47$ 
    & $67.24\pm 0.42$
    & $67.54^{+0.44}_{-0.50}$ 
    & $67.61^{+0.43}_{-0.48}$ 
    & $67.71\pm 0.44$ 
    & $67.44^{+0.50}_{-0.56}$ 
    & $67.40\pm 0.52$ \\
$r_d$ 
    & $147.80\pm 0.43$ 
    & $143.9^{+1.3}_{-0.98}$ 
    & $143.7^{+1.3}_{-1.1}$
    & $147.87\pm 0.43$
    & $143.5^{+1.3}_{-1.0}$
    & $143.3^{+1.2}_{-1.1}$
    & $147.59\pm 0.44$ 
    & $142.0^{+2.2}_{-1.7}$ 
    & $140.4\pm 2.4 $\\
$\Omega_b h^2$ 
    & $0.02237\pm 0.00013$ 
    & $0.02238\pm 0.00014$
    & $0.02238\pm 0.00013$
    & $0.02234\pm 0.00013$
    & $0.02236\pm 0.00014$ 
    & $0.02238\pm 0.00013$ 
    & $0.02242\pm 0.00013$ 
    & $0.02234\pm 0.00014$ 
    & $0.02234\pm 0.00014$ \\
$\Omega_{\phi}$ 
    & $0.6822\pm 0.0058$
    & $0.6853\pm 0.0064$ 
    & $0.6860\pm 0.0064$
    & $0.6805\pm 0.0058$
    & $0.6844\pm 0.0065$
    & $0.6853\pm 0.0063$
    & $0.6870\pm 0.0060$
    & $0.6830\pm 0.0073$
    & $0.6823\pm 0.0072$ \\
$w_{\phi}$ 
    & $-1.0$ 
    & $-0.763\pm 0.054$ 
    & $ -0.798\pm 0.046$
    & $-1.0$ 
    & $-0.734\pm 0.054$
    & $-0.776\pm 0.047$
    & $-1$ 
    & $-0.674^{+0.098}_{-0.063}$
    & $-0.64\pm 0.11$ \\
$\theta_{0}$ 
    & $-$ 
    & $1.340^{+0.062}_{-0.038}$ 
    & $1.350^{+0.065}_{-0.034}$
    & $-$ 
    & $1.325^{+0.063}_{-0.040}$
    & $1.338^{+0.066}_{-0.036}$
    & $-$ 
    & $1.302^{+0.038}_{-0.070}$
    & $1.296^{+0.034}_{-0.065}$ \\
$\beta$ 
    & $-$ 
    & $-0.81^{+0.36}_{-0.57}$ 
    & $-0.73^{+0.42}_{-0.63}$
    & $-$ 
    & $-0.81^{+0.41}_{-0.48}$
    & $-0.74^{+0.44}_{-0.62}$
    & $-$ 
    & $-0.83\pm 0.34$
    & $-0.91^{+0.26}_{-0.51}$ \\
$\alpha$ 
    & $-$ 
    & $4.4^{+2.1}_{-4.1}$ 
    & $-$ 
    & $-$ 
    & $4.4^{+2.1}_{-4.1}$
    & $-$ 
    & $-$ 
    & $3.2^{+1.2}_{-3.3}$
    & $-$ \\
$\lambda_{0}$ 
    & $-$ 
    & $2.45^{+0.59}_{-0.52}$ 
    & $1.74\pm 0.43$ 
    & $-$ 
    & $2.62^{+0.66}_{-0.41}$
    & $1.84\pm 0.44$
    & $-$ 
    & $2.81^{+0.68}_{-0.24}$
    & $2.42^{+0.71}_{-0.55}$ \\
$n$ 
    & $-$ 
    & $1.71^{+0.58}_{-1.1}$ 
    & $-$ 
    & $-$ 
    & $1.75^{+0.62}_{-1.1}$
    & $-$ 
    & $-$ 
    & $1.77^{+0.62}_{-1.1}$
    & $-$ \\
$m$ 
    & $-$ 
    & $-$ 
    & $3.0^{+1.6}_{-1.0}$
    & $-$ 
    & $-$
    & $3.1^{+1.6}_{-1.0}$
    & $-$ 
    & $-$
    & $2.9^{+1.0}_{-1.6}$ \\
$\chi_{min}^{2}$ 
    & $1794.038$
    & $1775.045$
    & $1774.823$ 
    & $1714.250$
    & $1690.449$
    & $1690.567$
    & $35.733$ 
    & $22.281$
    & $20.387$ \\
$\Delta \chi^{2}$ 
    & $0$
    & $-18.993$
    & $-19.215$ 
    & $0$
    & $-23.751$
    & $-23.683$
    & $0$ 
    & $-13.452$
    & $-15.346$ \\
$\Delta AIC$
    & $0$
    & $-8.993$
    & $-11.215$
    & $0$
    & $-13.751$
    & $-15.683$
    & $0$ 
    & $-3.452$
    & $-7.346$ \\
\bottomrule
\end{tabular}
}
\caption{Best-fit values and 68\% confidence intervals for key cosmological parameters in the $\Lambda$CDM, axion and inverse power law potentials for three data combinations.}
\label{tab:bestfit}
\end{table*}

Table-\ref{tab:bestfit}  provides the mean values and the associated $68\%$CL constraints on the cosmological parameters and model parameters from our analysis for each potential. The constraints reported here are based on the three combinations of data discussed in the previous section. We have also presented the constraint obtained on the $\Lambda$CDM model for comparison.

\begin{figure*}[!hbt]
            \centering
            \includegraphics[width=2\columnwidth]{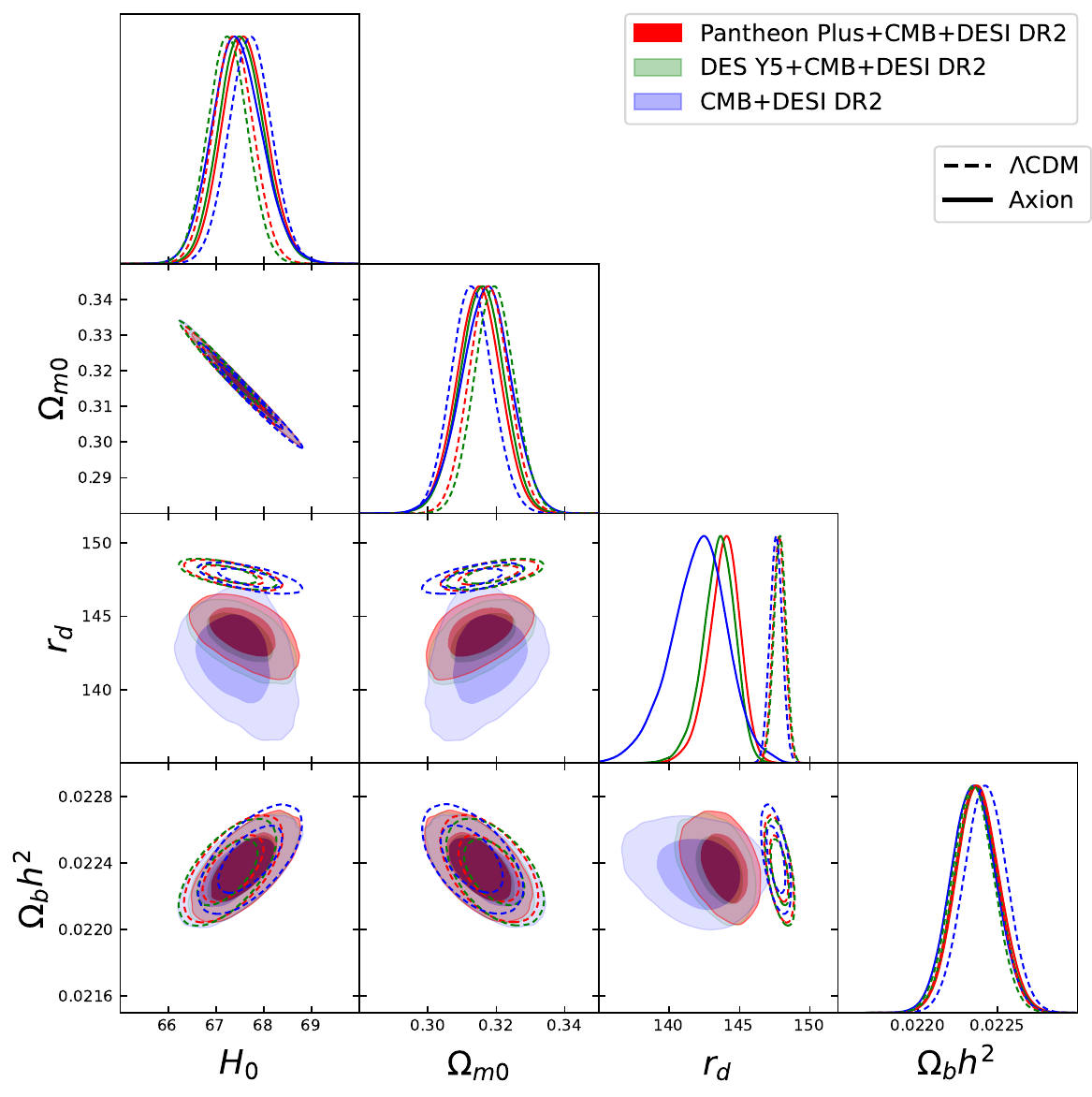}
            \caption{Plots of the cosmological parameters for the axion potential and a comparison with $\Lambda$CDM model.}
            \label{fig:Final_All_Axion_vs_lcdm}
    \end{figure*}

\begin{figure*}[!hbt]
            \centering
            \includegraphics[width=2\columnwidth]{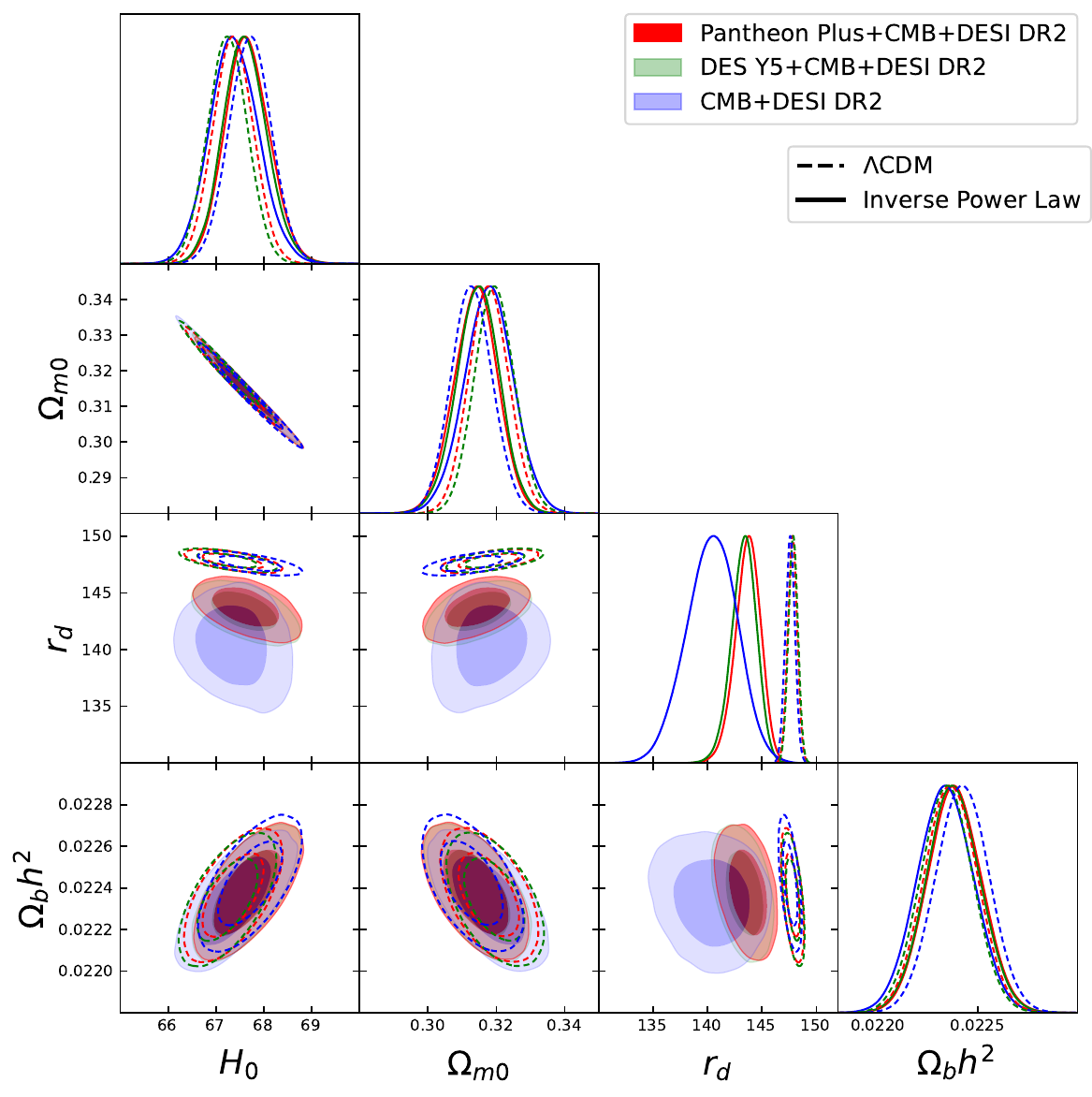}
            \caption{Plots of the cosmological parameters for the inverse power law potential and a comparison with $\Lambda$CDM model.}
            \label{fig:Final_All_InvPow_vs_lcdm}
    \end{figure*}

The $2D$ and $1D$ triangular plots depicting the cosmological parameters $H_{0}$, $\Omega_{m0}$, $r_{d}$, and $\Omega_{b}h^{2}$ are presented in FIG. \ref{fig:Final_All_Axion_vs_lcdm} for the axion potential and in Fig. \ref{fig:Final_All_InvPow_vs_lcdm} for the inverse power law potential. The plots for Set 1, Set 2, and Set 3 are depicted in red, green, and blue, respectively. For comparison of our models with the $\Lambda$CDM model, we have also plotted the results for the same cosmological parameters in a dotted line, whereas solid lines are used for both the axion and inverse power law potentials. From our analysis of both potentials, we found that the coupling parameter is nonzero and negative. We also find that $\beta$ is constrained from above without any lower bound.

$\bullet$ \textbf{Axion Potential}: For the axion potential, the additional parameters are $\theta_0$, $\beta$, $\alpha$, $\lambda_0$, and $n$ compare to $\Lambda$CDM model. For all three data combinations assessed, the matter density $\Omega_{m0}$ in the axion potential remains largely consistent with its $\Lambda$CDM counterpart, suggesting no notable discrepancies. The Hubble constant $H_0$ is slightly higher compared to $\Lambda$CDM for datasets Set 1 and Set 2, with central estimates ranging from $67.54$ to $67.60$ km/s/Mpc. But for the dataset Set 3, the value of $H_{0}$ is reduced compared to the $\Lambda$CDM model. It can be noticed here that the inclusion of Supernova data (in Set 1 and Set 2) increases the estimated mean value of $H_0$. The fitted value for the $r_d$ parameter for all three datasets is lower than that for the $\Lambda$CDM. The equation of state for dark energy, $w_\phi$ for the axion potential, deviates significantly from $-1$, with values between $-0.674$ and $-0.763$ for all three data sets, indicating a move away from a cosmological constant toward a quintessence-like behaviour.

    \begin{figure*}[!hbt]
            \centering
            \includegraphics[width=2\columnwidth]{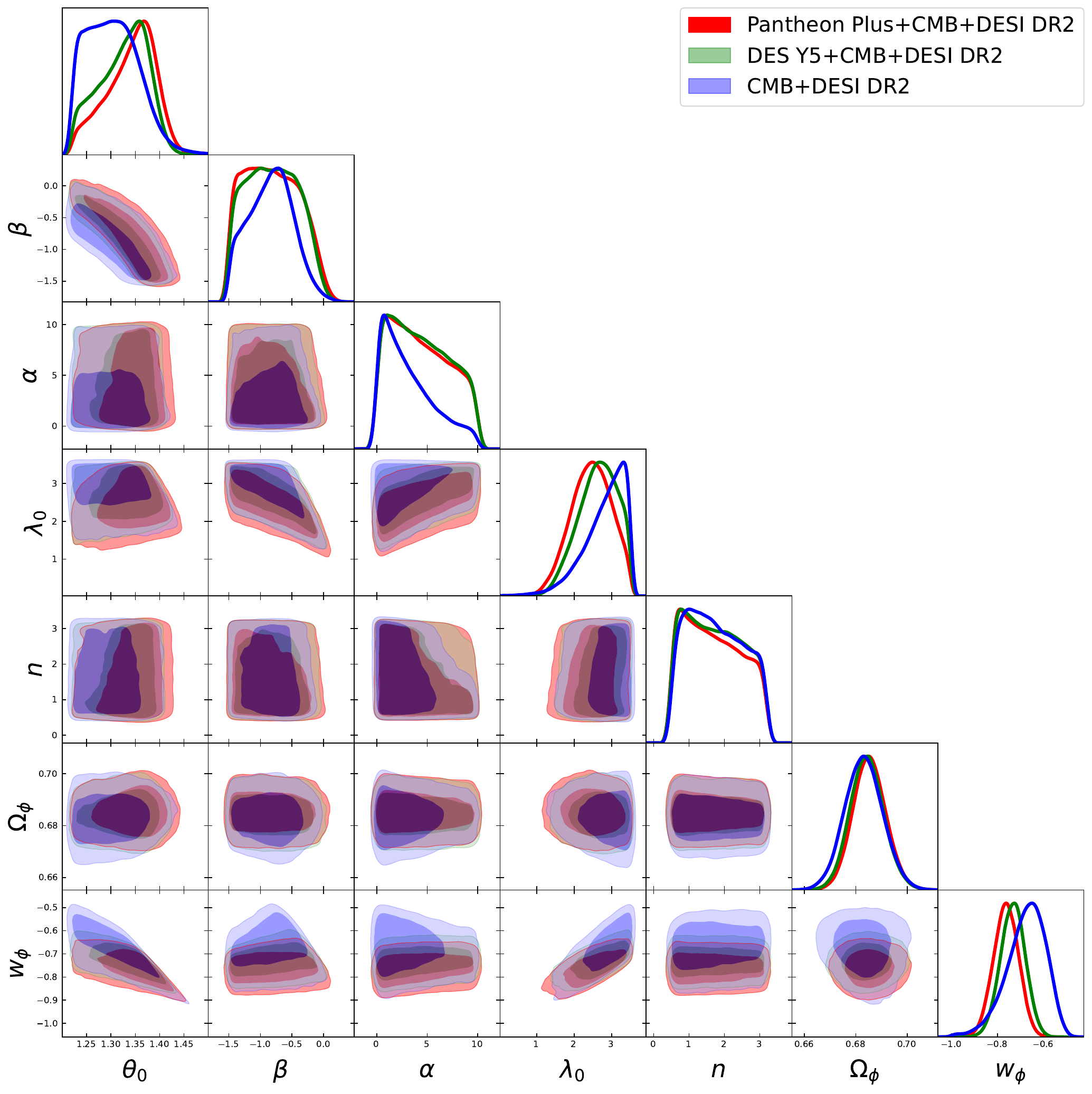}
            \caption{Plots of the cosmological parameters and model parameters for the axion potential.}
            \label{fig:Final_All_Axion}
    \end{figure*}

$\bullet$ \textbf{Inverse Power law potential:} For the inverse power law potential, additional model parameters are $\beta$, $\lambda_0$, and $m$. The mean value of the parameters $\Omega_{m0}$ and $H_0$ obtained for this potential are close to those found in both the $\Lambda$ CDM and the axion scenarios. However, for the inclusion of the Supernova data, $H_0$ exhibits a slight increase relative to $\Lambda$CDM for datasets Set 1 and Set 2, with central values between $67.61$ and $67.65$ km/s/Mpc. Like the axion potential, the value of $H_{0}$ is slightly reduced for the Set 3 data set, which includes only CMB and BAO data, compared to the $\Lambda$CDM model. Also, the mean value obtained for $r_d$ is lower than the $\Lambda$CDM. The scalar field's equation of state $w_\phi$ deviates from $-1$, reaching values up to $-0.64$ in Set 3. For Set 1 and Set 2, the $w_\phi$ is lower for the InvPow potential than the axion potential. Whereas for Set 3, it is reversed.

    \begin{figure*}[!hbt]
            \centering
            \includegraphics[width=2\columnwidth]{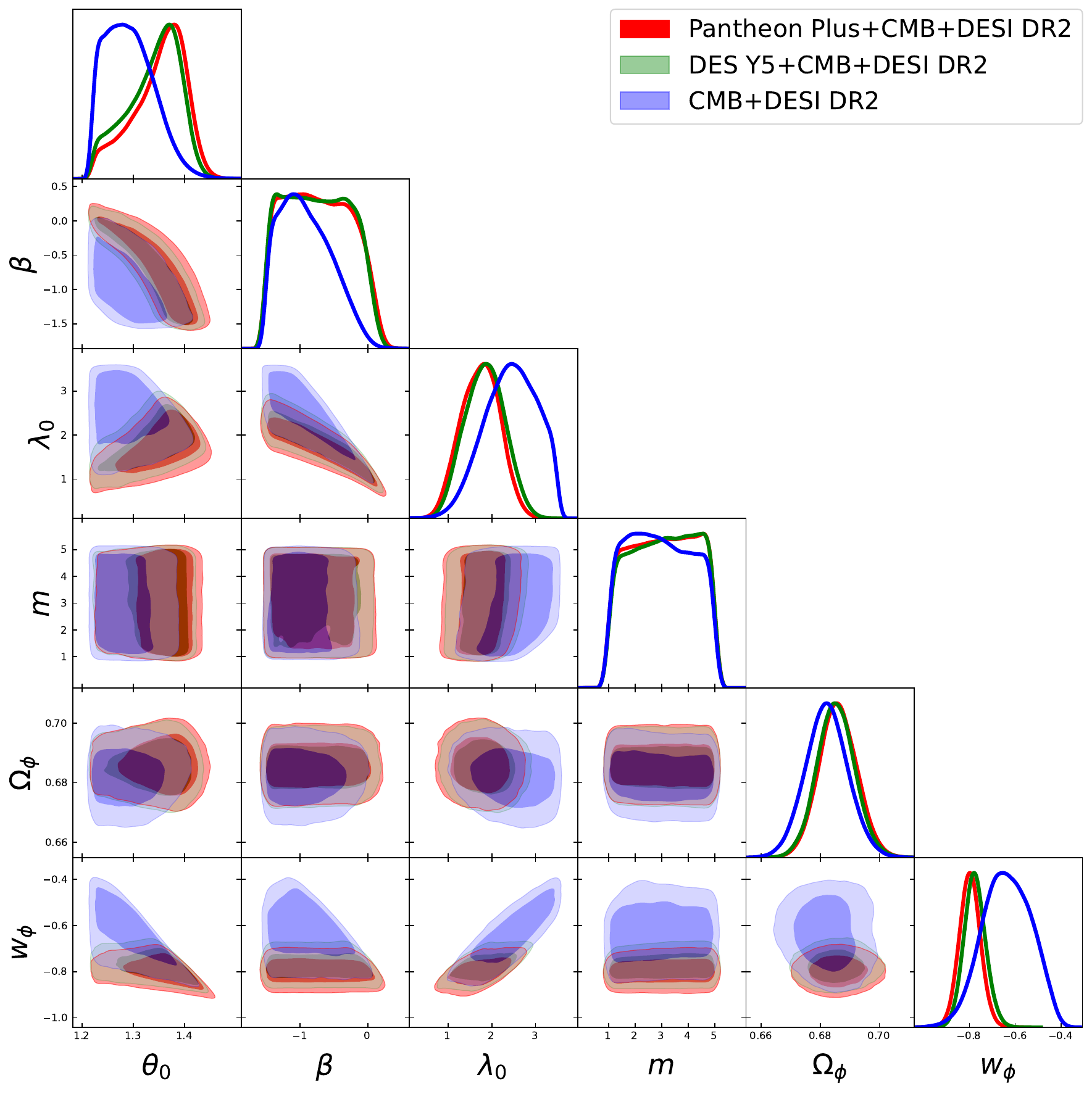}
            \caption{Plots of the cosmological parameters and model parameters for the inverse power law potential.}
            \label{fig:Final_All_InvPow}
    \end{figure*}

\section{ Model Selection}

In this section, we discussed the relative statistical performance of our model against the $\Lambda$CDM model for both the potentials. The comparison is based on the minimum chi-squared value ($\chi^2_{\text{min}}$), the difference in chi-squared ($\Delta \chi^2$), and the difference in Akaike Information Criterion ($\Delta \text{AIC}$). We have computed the difference in minimum chi-square values, $\Delta\chi^2_{\text{min}}$, for the axion potential and inverse power law potential relative to the $\Lambda$CDM model, as presented in Table-\ref{tab:bestfit}. For each dataset, the axion potential and the inverse power law potential fit the observational data significantly better than the $\Lambda$CDM model.  For Set 1, the inverse power law potential yields the lowest value of $\Delta \chi^2_{\text{min}} = -19.165$, suggesting a significantly better fit to the observational data compared to the other two models. For Set 2 axion potential fit the data better with $\Delta \chi^2_{\text{min}}=-23.836$, whereas for Set 3 inverse power law potential better fit the data with $\Delta \chi^2_{\text{min}}=-15.322$.

The difference of AIC is defined as: 

\begin{equation}\label{Friedmann_constraint_polar}
 \Delta AIC = \chi^{2}_{min, \mathcal{M}} - \chi^{2}_{min, \Lambda CDM} + 2(N_{\mathcal{M}} - N_{\Lambda CDM}),
\end{equation}

Here, $\mathcal{M}$ represents the model, and $N_{\mathcal{M}}$ and $N_{\Lambda CDM}$ are the parameters for model $\mathcal{M}$ and for $\Lambda CDM$, respectively. Generally, if $\Delta AIC$ is negative compared to the $\Lambda CDM$ model, it signals a preference for the model $\mathcal{M}$. To evaluate the models with consideration of their complexity, we analyze the  Akaike Information Criterion (AIC), which penalizes models with additional parameters. Following the standard interpretation of AIC differences, the values between -4 and -2 indicate evidence for the weak and -7 to -4 indicate moderate support, while the values less than -10 suggest strong support.

When using data Set1, Set 2 and Set 3, the $\Delta AIC = -11.215$, $\Delta AIC = -15.683$ and $\Delta AIC = -7.346$ for inverse power law potential respectively. Whereas for the same data sets, $\Delta AIC = -8.993$, $\Delta AIC = -13.751$ and $\Delta AIC = -3.452$ for the axion potential, respectively. This indicates that the inverse power-law potential is more preferable than the axion potential over the $\Lambda$CDM for all the data sets. For Set 1 and Set 2 inverse powerlaw potential shows strong evidence for the momentum-coupled model over the  $\Lambda$CDM. For Set 3 data is reduced to the moderate evidence. For the axion potential, Set 1 shows moderate evidence, Set 2 shows strong evidence, and Set 3 shows weak evidence.

\section{Evolution of Hubble and density parameters}
\begin{figure}[!hbt]
            \centering
            \includegraphics[width=\columnwidth]{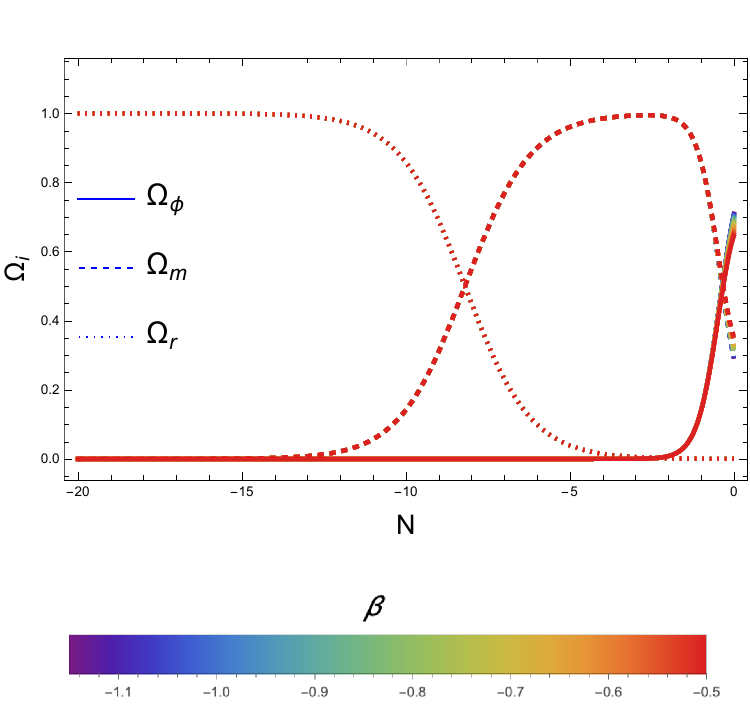}
            \caption{Evolution of density parameters for axion potential with the range of $\beta : [-1.15, -0.5]$. The dotted line shows the evolution of the $\Omega_r$, the dashed line shows the evolution of the $\Omega_m$, and the solid line corresponds to the evolution of $\Omega_\phi$.}
            \label{fig:axiondensparamPlot1}
    \end{figure}

\begin{figure}[!hbt]
            \centering
            \includegraphics[width=\columnwidth]{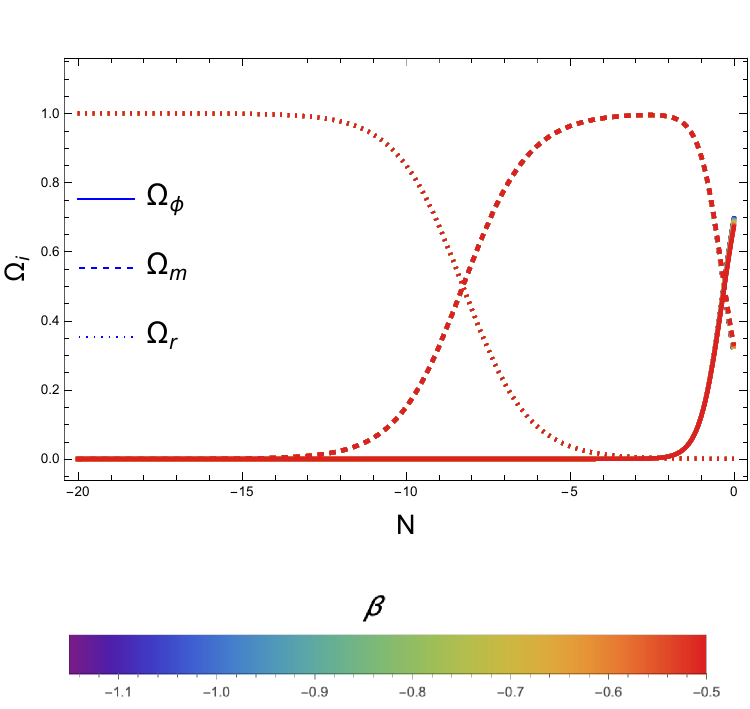}
            \caption{Evolution of density parameters for inverse power law potential with the range of $\beta : [-1.15, -0.5]$. The dotted line shows the evolution of the $\Omega_r$, the dashed line shows the evolution of the $\Omega_m$, and the solid line corresponds to the evolution of $\Omega_\phi$.}
            \label{fig:invpowdensparamPlot1}
    \end{figure}

\begin{figure}[!hbt]
            \centering
            \includegraphics[width=\columnwidth]{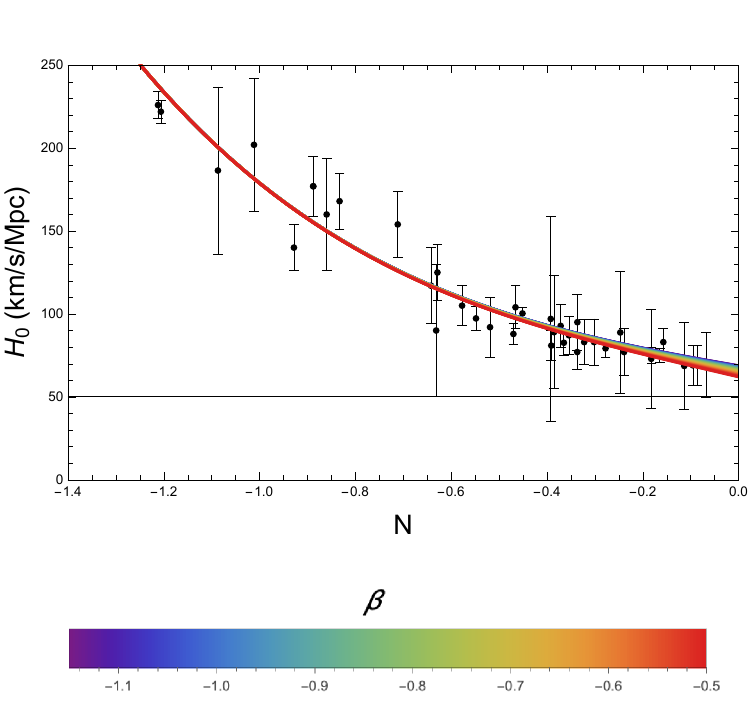}
            \caption{Evolution of Hubble parameter for axion potential with the range of $\beta : [-1.15, -0.5]$. The observational data have been plotted for comparison.}
            \label{fig:HubbleAxionEvol}
    \end{figure}

\begin{figure}[!hbt]
            \centering
            \includegraphics[width=\columnwidth]{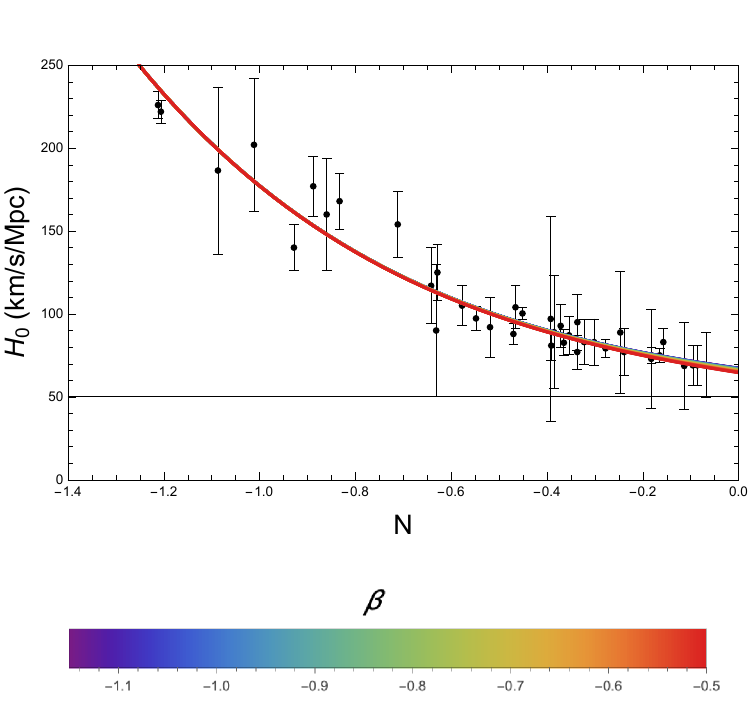}
            \caption{Evolution of Hubble parameter for inverse power law potential with the range of $\beta : [-1.15, -0.5]$. The observational data have been plotted for comparison.}
            \label{fig:Hubbleevol_InvPow}
    \end{figure}

To explore the dynamics of our universe within a scalar field framework featuring axion and inverse power-law potentials separately, we solve Eq.(\ref{axion_autonomous:1_momentum}) and Eq.(\ref{invpow_autonomous:1_momentum}) numerically. This is for the evolution of the Hubble parameter $H(N)$ and the density parameters $\Omega_i(N)$ with respect to $N$. The analysis includes radiation to the total energy budget of the universe to ensure that the Friedmann constraint holds throughout the evolution of these density parameters. This analysis is key to examining the cosmic expansion and energy composition at varying redshifts. We numerically solve the equations for a range of $\beta : [-1.15, -0.5]$, which is obtained from our MCMC analysis and includes all the posteriors for the three different sets. For the axion potential, initial conditions are set at $N=-12.9$, with $\alpha = 4.4$, $n = 1.75$, $r(-12.9) = \sqrt{5 \times 10^{-19}}$, $\theta(-12.9) = 1.33$, $\Omega_m(-12.9) = 0.009 - r(-12.9)$, and $\lambda(-12.9) = 1.67$. For the inverse power-law potential the initial conditions are taken at $N=-13.08$, with $m = 3$, $r(-13.08) = \sqrt{2.2 \times 10^{-19}}$, $\theta(-13.08) = 1.34$, $\Omega_m(-13.08) = 0.008 - r(-13.08)$, and $\lambda(-13.08) = 1.8$. FIG. \ref{fig:axiondensparamPlot1} and FIG. \ref{fig:invpowdensparamPlot1} illustrates the evolution of density parameters for the axion and inverse power law potential respectively, validating our parameter and initial condition selections, which indicates that the Universe progresses from a radiation state to a matter-dominated state, and in recent time, the universe is dominated by the scalar field. These findings demonstrate both scalar field models' capacity to mimic the Universe's background expansion, transitioning from radiation to matter, and ultimately to dark energy dominance. FIG. \ref{fig:HubbleAxionEvol} and FIG. \ref{fig:Hubbleevol_InvPow} display the evolution of the Hubble parameter with the observed data from cosmic chronometer data set (see Table VII of Ref.\cite{Sahoo:2025vtu} ) for the axion and inverse power law potentials, respectively, showing alignment with observed values at late time.  Moreover, these behaviors are stable under small variations in initial conditions, except for $\lambda$ and robust to the chosen range of the coupling parameter $\beta$, showcasing the viability of these models in explaining cosmic acceleration.

\section{Stability Analysis}
In this section, we discuss the system's stability for both the potentials mentioned earlier. To do that from equations Eq.(\ref{axion_autonomous:1_momentum}) and Eq.(\ref{invpow_autonomous:1_momentum}), the fixed points (or equilibrium points) $\left(r_{c}, \theta_{c}, \lambda_{c}\right)$ are obtained by solving the system of equations $r^{'}=0$, $\theta^{'}=0$ and $\lambda^{'}=0$ simultaneously. The list of fixed points, their stability, and the values of cosmological parameters for the system of equations Eq.(\ref{axion_autonomous:1_momentum}) (axion potential) and Eq.(\ref{invpow_autonomous:1_momentum}) (inverse power law potential) are given in Table-\ref{tab:fixedpoints_Axion_momentum} and Table-\ref{tab:fixedpoints_invpow_momentum}, respectively. The fixed points for the system of equations Eq.(\ref{axion_autonomous:1_momentum}) corresponding to the axion potential are denoted by $P_{i}$ (i=1,2,3) and those for the system of equations Eq.(\ref{invpow_autonomous:1_momentum}) corresponding to the inverse power law potential are denoted by $Q_{j}$ (j=1,2,3,4). One can study the stability of a fixed point by finding out the eigenvalues of the corresponding Jacobian matrix at that fixed point. If all the eigenvalues have negative real parts, then the fixed point is stable. However, if at least one eigenvalue with a positive real part exists, then the fixed point is unstable. 

\begin{table*}[!hbt]
    \centering
    \scalebox{1.0}{
    \begin{tabular}{|c|c|c|c|c|c|c|c|c|}
       \hline {\vtop{\hbox{\strut Equilibrium}\hbox{\strut points}}} & { $r_{c}$} &{$\theta_{c}$} & {$\lambda_{c}$}&   {Eigenvalues} & {Stability} & {$q$} & {$w_{\phi}$} & {$w_{tot}$} \\ \hline
        $P_{1}$ & $0$  &  $n\pi$ & any  & $-\frac{3}{2}$, $0$, $3$ & saddle point  &  $\frac{1}{2}$ &  $1$  &  $0$  \\  \hline
        $P_{2}$ & $0$  &  $2n\pi \pm \frac{\pi}{2}$ & any  & $-3$, $0$, $\frac{3}{2}$ & saddle point  &  $\frac{1}{2}$ &  $-1$  &  $0$  \\  \hline
        $P_{3}$ & $1$  &  $2n\pi \pm\frac{\pi}{2}$ & $0$  & $-3$, $-\frac{3}{2}\pm \frac{\sqrt{9+6n\alpha -18\beta}}{2\sqrt{1-2\beta}}$ & \vtop{\hbox{\strut stable node for }\hbox{\strut $6\beta > 3 + 2n\alpha$}}  &  $-1$ &  $-1$  &  $-1$  \\  \hline
         \end{tabular}
         }
    \caption{ The list of fixed points and their stability for the system of equations given in Eq.$(\ref{axion_autonomous:1_momentum})$.}
    \label{tab:fixedpoints_Axion_momentum}
\end{table*}

\begin{table*}[!hbt]
    \centering
    \scalebox{1.0}{
    \begin{tabular}{|c|c|c|c|c|c|c|c|c|}
       \hline {\vtop{\hbox{\strut Equilibrium}\hbox{\strut points}}} & { $r_{c}$} &{$\theta_{c}$} & {$\lambda_{c}$}&   {Eigenvalues} & {Stability} & {$q$} & {$w_{\phi}$} & {$w_{tot}$} \\ \hline
        $Q_{1}$ & $0$  &  $n\pi$ & any  & $-\frac{3}{2}$, $0$, $3$ & saddle point  &  $\frac{1}{2}$ &  $1$  &  $0$  \\  \hline
        $Q_{2}$ & $0$  &  $2n\pi \pm \frac{\pi}{2}$ & any  & $-3$, $0$, $\frac{3}{2}$ & saddle point  &  $\frac{1}{2}$ &  $-1$  &  $0$  \\  \hline
        $Q_{3}$ & $\frac{1}{\sqrt{1-2\beta}}$  &  $n\pi $ & $0$  & $0$, $3$, $3$ & unstable node  &  $2$ &  $1$  &  $1$  \\  \hline
        $Q_{4}$ & $1$  &  $2n\pi \pm\frac{\pi}{2}$ & $0$  & $-3$, $-3$, $0$ & stable node  &  $-1$ &  $-1$  &  $-1$  \\  \hline
        
         \end{tabular}
         }
    \caption{ The list of fixed points and their stability for the system of equations given in Eq.$(\ref{invpow_autonomous:1_momentum})$.}
    \label{tab:fixedpoints_invpow_momentum}
\end{table*}

\begin{figure*}[!hbt]
            \centering
            \includegraphics[width=2\columnwidth]{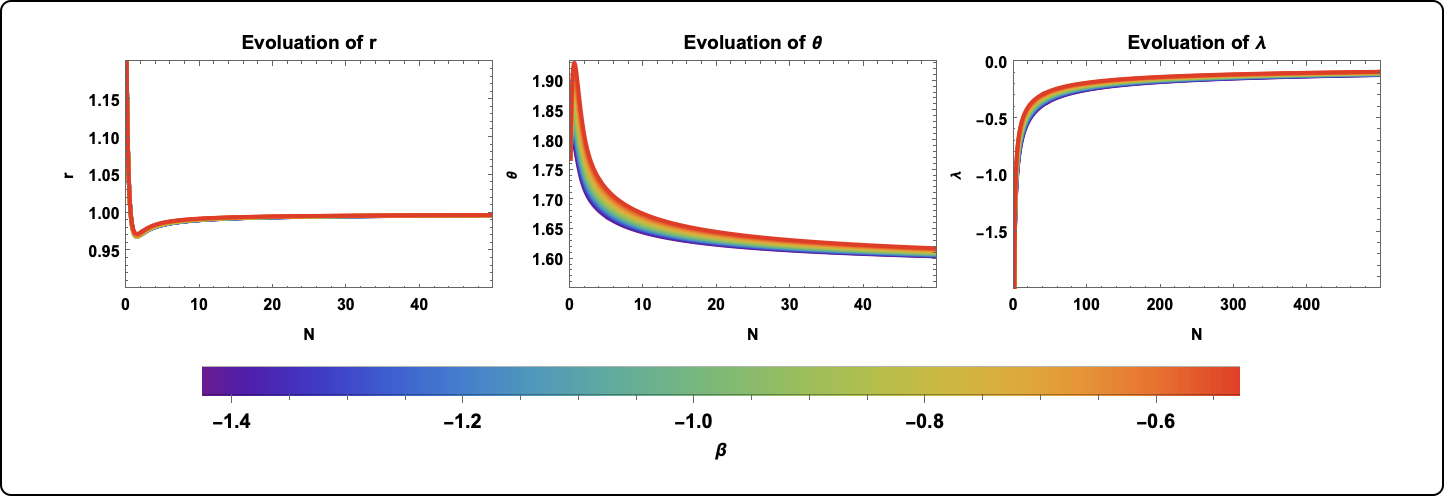}
            \caption{Plots of the numerical solutions for each dynamical variable for the system of equations Eq.(\ref{invpow_autonomous:1_momentum}) by considering perturbations around the equilibrium point $Q_{4}$ with a range of values of $\beta :[-1.15, -0.5]$ and $m=3$.}
            \label{fig:Stability_InvPow_Momentum}
    \end{figure*}

By using this technique, one can find that for the axion potential $P_1$ and $P_2$ are saddle, whereas $P_3$ can be stable. The fixed point $P_1$ and $P_2$ correspond to the matter dominated era which represents a decelerating universe. At the fixed point $P_1$ the EoS of the scalar field is $w_\phi=1$ hence the scalar field behaves as a stiff fluid. On the other hand, at the fixed point $P_2$ the EoS of the scalar field is $w_\phi=-1$, representing a cosmological constant like scenario.  The fixed point $P_{3}$ is isolated as well as a hyperbolic fixed point. The stability of $P_{3}$ depends on the parameters $n$, $\alpha$ and $\beta$. As $\beta<1/2$ for a physically viable model, we must have $6\beta > 3 + 2n\alpha$, otherwise there will be a positive eigenvalue of the Jacobian matrix for the fixed point $P_{3}$. For $6\beta > 3 + 2n\alpha$, there is one negative real and two complex eigenvalues with negative real part of the Jacobian matrix for the system of equations Eq.(\ref{axion_autonomous:1_momentum}) at $P_{3}$. Thus, the fixed point $P_{3}$ is a stable spiral node for $6\beta > 3 + 2n\alpha$. This indicates that the fixed points given by $P_{3}$ are locally asymptotically stable. This fixed point is completely dark energy dominated and represents an accelerating universe with $w_\phi = w_{tot} =-1$.

In the context of the inverse power law potential, the fixed points $Q_1$ and $Q_2$ exhibit saddle-like characteristics, indicating a matter-dominated, decelerated universe. Analogous to the situation with axions, at the fixed point $Q_1$, the scalar field behaves as a stiff fluid characterized by $w_\phi=1$. In contrast, at the fixed point $Q_2$, the equation of state of the scalar field is $w_\phi=-1$, which suggests a scenario similar to a cosmological constant. Fixed point $Q_3$ represents a complete scalar field dominated decelerated universe in which the scalar field behaves like a stiff fluid. This fixed point is unstable in nature.

The eigenvalues of the Jacobian matrix for the fixed point $Q_{4}$ has a mixture of negative and zero real parts. Due to the presence of a zero real part, the stability of $Q_{4}$ can not be obtained by using the linear stability analysis. In this scenario, one can either apply analytical technique like  center manifold theory, or numerical investigation like the evaluation of the dynamical variables by taking a perturbation around the fixed point $Q_{4}$. Here we adopt the last one. In this method, we perturb the system around the $Q_{4}$ fixed point and plotted the evolution of the system projected on the each dynamical variable.  The evaluation of all dynamical variables with respect to $N$ for the system of equations Eq.(\ref{invpow_autonomous:1_momentum}) are given in fig.(\ref{fig:Stability_InvPow_Momentum}). One can see from these plots that the evolution of the system approaches the $Q_{4}$ fixed point after the perturbation. From this analysis, we can say the fixed point given by $Q_{3}$ can be a stable one for $\beta$ ranges between $-1.42$ and $-0.48$. The range of $\beta$ considered here includes the posterior of the $\beta$ parameter obtained from the MCMC analysis for the inverse power law potential (please see Table:\ref{tab:bestfit}).

\section{Conclusion}

In this study, we developed a dynamical dark energy model within a spatially flat FLRW universe, introducing a momentum transfer interaction between the scalar field and matter. The interaction term, $\mathcal{L}_{\text{int}} = -\beta (u^\alpha \partial\alpha \phi)^2$, modifies the scalar field dynamics and, consequently, the cosmic expansion history. We derived the modified background equations, including the generalized Klein-Gordon equation, and obtained expressions for the energy density and pressure of the scalar field. Physical consistency requires $\beta < \frac{1}{2}$, as larger values lead to ghost instabilities.

To explore the cosmological evolution, we reformulated the system into autonomous differential equations via a polar transformation of the phase space variables. This allowed a compact and unified analysis of the background dynamics. In this work, we focused on two theoretically motivated potentials: the axion potential and the inverse power-law potential. 

We carried out an MCMC analysis using three combinations of current cosmological datasets: Pantheon Plus, DES Y5, DESI DR2, and Planck18 compressed likelihoods. Both the axion and inverse power law potentials were found to show moderate to strong evidence over the $\Lambda$CDM. Inclusion of the Supernova data seems to show stronger evidence for the dynamical dark energy. The inverse power law potential shows more preference over the axion potential. Interestingly, the coupling constant $\beta$ is consistently constrained to negative values, with no evident lower bound, suggesting a momentum-coupled interaction between the dark sectors. From our dynamical system analysis, we found that the complete dark energy domination is the late-time attractor of the universe. Interestingly, the scalar field can behave as a stiff fluid at early times.

In summary, both the axion and inverse power law scalar field models, constructed within a momentum-transfer interaction framework, provide compelling alternatives to $\Lambda$CDM. Supported by both observational data and dynamical system analysis.



\appendix

\section{Prior on model parameters}\label{Prior on model parameters}
For both potentials, the EoS of the dark energy depends on  $\theta$ and the coupling parameter $\beta$ through Eq. (\ref{EoS_quintessence}). To determine the prior of these two parameters, such that $w_{\phi}\sim -1\pm 0.3$, we followed the following approximation. From our definition of the $w_\phi$ when $\theta$ is an odd multiple of $\pi/2$,  $w_{\phi}=-1$. To obtain the variation in $w_{\phi}$ around -1 value one can consider a small variation $\delta$ of $\theta$ around any odd multiple of $\pi/2$, that is, $\theta \sim (2n+1)\frac{\pi}{2}+\delta$. Also when $\cos \theta \sim \pm \delta$ we can approximate $w_{\phi}\sim \frac{2(1-\beta)\delta^{2}-1}{-2\beta \delta^{2}+1}$. The range of $\delta$ can be obtained from this approximated $w_{\phi}$, depending on whether $\beta$ is positive or negative. For $\beta<0$ we have $\left| \delta \right|\leq \sqrt{\frac{0.03}{2-4.1\beta}}$. Again for $0 \leq\beta <1/2$, it gives $\left| \delta \right|\leq \sqrt{\frac{0.03}{2-3.94\beta}}$ with $0<\beta<20/41$. Hence for the choice of $\beta :[-1.5, 0.48]$, the prior on $\theta$ comes out to be $\theta \sim [1.22, 1.91]$ which ensure that $w_{\phi}\sim -1\pm 0.3$.

\clearpage
\onecolumngrid
\twocolumngrid 

\bibliographystyle{unsrt}
\bibliography{sample}

\end{document}